# Estimating Network Parameters for Selecting Community Detection Algorithms

Leto Peel
Advanced Technology Centre,
BAE Systems,
Bristol, UK
leto.peel@baesystems.com

***Abstract** – This paper considers the problem of algorithm selection for community detection. The aim of community detection is to identify sets of nodes in a network which are more interconnected relative to their connectivity to the rest of the network. A large number of algorithms have been developed to tackle this problem, but as with any machine learning task there is no "one-size-fits-all" and each algorithm excels in a specific part of the problem space. This paper examines the performance of algorithms developed for weighted networks against those using unweighted networks for different parts of the problem space (parameterised by the intra/inter community links). It is then demonstrated how the choice of algorithm (weighted/unweighted) can be made based only on the observed network.*

## 1 Introduction

The study of large scale networks has revealed a number of properties about the behaviour and topology of naturally occurring networks. One such property is the presence of community structures; sets of nodes in a network which are more interconnected relative to their connections to the rest of the network. The aim of community detection is to identify these structures. Community detection is a problem which has attracted much interest in recent years [1-5] and has consequently produced a wide range of approaches to the problem; an in-depth review of most contemporary methods is given in [6].

One of the reasons why the ability to detect communities is so attractive lies in the phenomenon known as *assortative mixing*, where entities in a network are observed to associate preferentially with similar entities. This suggests that detecting communities may be used for identifying entities which share common attributes or purposes. An example of community structures corresponding to entity similarity is given in [7] where community structures in a friendship network identify similarities in race and age. The wide range of complex systems that can naturally be expressed as networks (human interaction patterns, metabolic networks, WWW, the brain) implies that community detection has applications spanning domains as diverse as biology [8-10], sociology [8],[11],[12], computer science [13],[14] and intelligence [15-17].

The implications of community detection in the intelligence domain are that it could be used to identify groups of people who share common goals or purposes. To this effect, community detection could potentially be used to constrain the inference problem when investigating or detecting malicious activities, e.g. rather than monitoring all people, use community detection as a pre-processing step to select a subset of people to monitor. In this setting, the network nodes would represent people and the links would represent interactions or relationships between them; such a network can be constructed from a database of phone records, email logs or other transactional data.

With a large selection of algorithms available to undertake the task of community detection, choosing an appropriate algorithm becomes problematic. This is largely due to the lack of formal or commonly accepted evaluation procedures. The networks used to evaluate community detection tend to be a small selection of real networks and/or networks generated from simple models, where these networks vary widely between authors. Recent work to address this has focused on developing benchmark networks [18] on which comparative analysis [19] can be drawn to determine the reliability of different algorithms. However, it is commonly accepted across the machine learning community that there is no *one-size-fits-all* solution and so this work considers the idea that for different situations, different classes of algorithms may outperform other classes of algorithms. The range of community detection algorithms in itself poses the intelligence analyst with the challenge of choosing an appropriate solution or combinations of solution techniques for the specific problem at hand. It is therefore desirable to be able to provide the intelligence analyst, who will likely not have expert knowledge of these algorithms, with appropriate guidance. This paper considers the problem of automatically selecting community detection algorithms based on observations of the community structure.

It has been previously observed how structural properties of communities affect the performance of community detection algorithms [19]. These properties cannot be measured from the network data alone as they require knowledge of the underlying community assignment. The main contribution of this work is to demonstrate how these structural properties can be estimated from features of the observed network. Therefore a prediction about which algorithm will perform best can be made. This is achieved by considering algorithms for weighted networks and algorithms for unweighted networks as two separate classes and demonstrating how the performance of these two classes differs across the problem space (defined in section 2). Finally, a Support Vector Machine (SVM) [20] is used to classify the networks according to the algorithm which will perform best.

The rest of the paper is organised as follows: Section 2 defines the problem space by defining the network and community structure types and the target algorithm classes. The performance of the algorithm classes with respect to the structural parameters is evaluated in Section 3. Section 4 describes the *observable* network parameters and how a mapping can be made from these to the underlying structural parameters. The results of using the observable parameters to choose an appropriate class of algorithm are given in Section 5. Conclusions are given in Section 6.

## 2  Problem Space

A network is a structure made up of nodes, representing entities, and links or edges, representing relationships or interactions between entities. The total number of links connected to a node is known as its degree. The network links may also have weights associated with them which may represent the relative importance of the link. For example, in an interaction network representing a phone record database, the nodes would represent people and the links phone calls. The link weights could then represent the frequency of calls. Network links may also be directed, but this will not be considered in this work.

The premise of community detection is that there is some underlying assignment of nodes to communities which has to be discovered. But despite the large amount of literature on the subject there is still a lack of agreement on what defines a community beyond the intuitive concept that community structures have more intra-community links than inter-community links. Without a common definition it is difficult to draw a comparison between algorithms. However, it may not be necessary (or even desirable) to define a specific common definition of community, as definitions may be dependent on the application. Instead perhaps a suitably comprehensive parameter set for describing the space of community types and structures of interest.

A reasonable starting point is the parameter set used to generate networks and communities using the Lancichinetti-Fortunato-Radicchi (LFR) benchmark generator [18] as not only do these describe a number of network properties, but by using the generator it is possible to obtain networks and community assignments with those properties. This parameter set is described in section 2.1.

### 2.1  Network-Community Parameterisation

The parameter set used to describe the problem space are the parameters used by the LFR benchmark which is fully described in [18]. The LFR benchmark was designed to generate datasets to test community detection algorithms and mimic the observed properties of large-scale real complex networks [21], such as power-law degree and community distribution.

The parameters are best described in the context of the graph generation procedure:

1. *N* nodes are assigned to communities such that the community size distribution conforms to a power-law with minus exponent $\tau_2$.

2. Each node is assigned a degree such that the degree distribution conforms to a power law with minus exponent $\tau_1$ and mean degree *k*.

3. Links are initially assigned randomly according to the degree distribution. A topological mixing parameter, $\mu_t$, is set to define the proportion of each nodes links which link outside its community. Topological consistency with this parameter is achieved through an iterative re-wiring procedure.

4. Each node is then assigned a strength according to a power-law distribution with minus exponent $\beta$. The strength of a node is the weighted analogy of degree and as such represents the sum of the weights of the links for a given node.

5. To assign the link weights a similar process to step 3 is carried out according to the weight mixing parameter, $\mu_w$.

It is accepted that these may not be a full set of parameters to comprehensively describe the space of all possible network-community structures. Even so, the space is one of high dimensionality and so full exploration of all the parameters is beyond the scope of this paper and remains for future work. To constrain the problem, the values of all parameters were fixed with the exception of $\mu t$ and $\mu w$, which from initial tests were found to have the greatest impact on use of link weights.

## 2.2 Algorithm Overview

The algorithm selection problem has been constrained to choosing between the class of algorithms which use link weight information and the class that does not. In light of this, it was decided to use algorithms suitable for unweighted or weighted networks. This way a controlled comparison can be drawn between the performances of the unweighted and weighted algorithms without needing to worry about differences in algorithms. Two such algorithms are examined:

- *Infomap* [5]: This algorithm approaches the community detection problem by identifying a duality between community detection and information compression. By using random walks to analyse the information flow through a network it identifies communities as modules through which information flows quickly and easily. Coding theory is used to compress the data stream describing the random walks by assigning frequently visited nodes a shorter codeword. This is further optimised by assigning unique codewords to network modules and reusing short codewords for network nodes such that node names are unique given the context of the module. This two level description of the path allows a more efficient compression by capitalising on the fact that a random walker spends more time within a community than moving between communities.
- *COPRA* [22]: This is an extension of the label propagation based RAK algorithm [23]. The algorithm works as follows; to start, all nodes are initialised with a unique label. These labels are then updated iteratively, where a node's new label is assigned according to the label used most by its neighbours. If there is more than one most frequently occurring label amongst the neighbours, then the label is chosen randomly. At termination of the algorithm, nodes with the same label are assigned to the same community. The **C**ommunity **O**verlap **PR**opagation **A**lgorithm (COPRA) extends the RAK algorithm to deal with the possibility of overlapping communities (although this aspect of community detection is not explored within this work). This is done by augmenting the label with a belonging factor such that for a given node these sum to 1. To prevent all nodes becoming a member of all communities, a threshold is set below which the labels are discarded. Due to the stochastic nature of the algorithm, particularly in the initial iterations, the algorithm is, in practise, run a number of times and the "best" community assignment is decided according to the one which has the highest modularity [24]. In the weighted instance of the algorithm, the weights of the network are incorporated by weighting the frequency of the labels according to the link weight connecting the respective node.

# 3 Algorithm Performance

A number of different metrics are used in the literature to measure the performance of community detection algorithms, however the Normalised Mutual Information [1] metric is one which has become fairly standard recently and so will be used here. This metric provides a measure of similarity between the algorithm output assignment and the true community assignment, where a value of 1 denotes a perfect match. Experiments were run to examine the effect of varying the two mixing parameters $\mu t$ and $\mu w$, the results of which can be seen in Figure 1.

Figure 1 shows the mutual information scores for the weighted algorithms (COPRAw, INFOMAPw) and unweighted algorithms (COPRAuw, INFOMAPuw) as $\mu w$ is changed. The plots (a) – (d) show the performance for different values of $\mu t$. Each point on the graphs represents the average mutual information over 25 generated networks with the indicated parameter values. It can be seen that the unweighted algorithms perform well when $\mu t$ is low and are unaffected by $\mu w$ for all values $\mu t$. This is only to be expected as these algorithms only rely on the topological information. The weighted algorithms on the other hand are affected by both parameters, but are seen to consistently perform well for low $\mu w$. The effect of $\mu t$ is probably best observed in Figure 2. Here it can be seen that the weighted algorithms perform well when $\mu t$ is at least as high as $\mu w$ (in this case $\mu w$=0.3). A similar observation was made in [19] where it was seen that weighted algorithms performed better overall at $\mu t$ values of 0.5, in comparison to lower values. It was explained that the reason for this is that a low $\mu t$ relative to $\mu w$ means that there is a lower proportion of inter-community links relative to the proportion of inter-community weights. The effect of this is that a small number of inter-community links receive high link weights relative to the intra-community weights, see Figure 3.

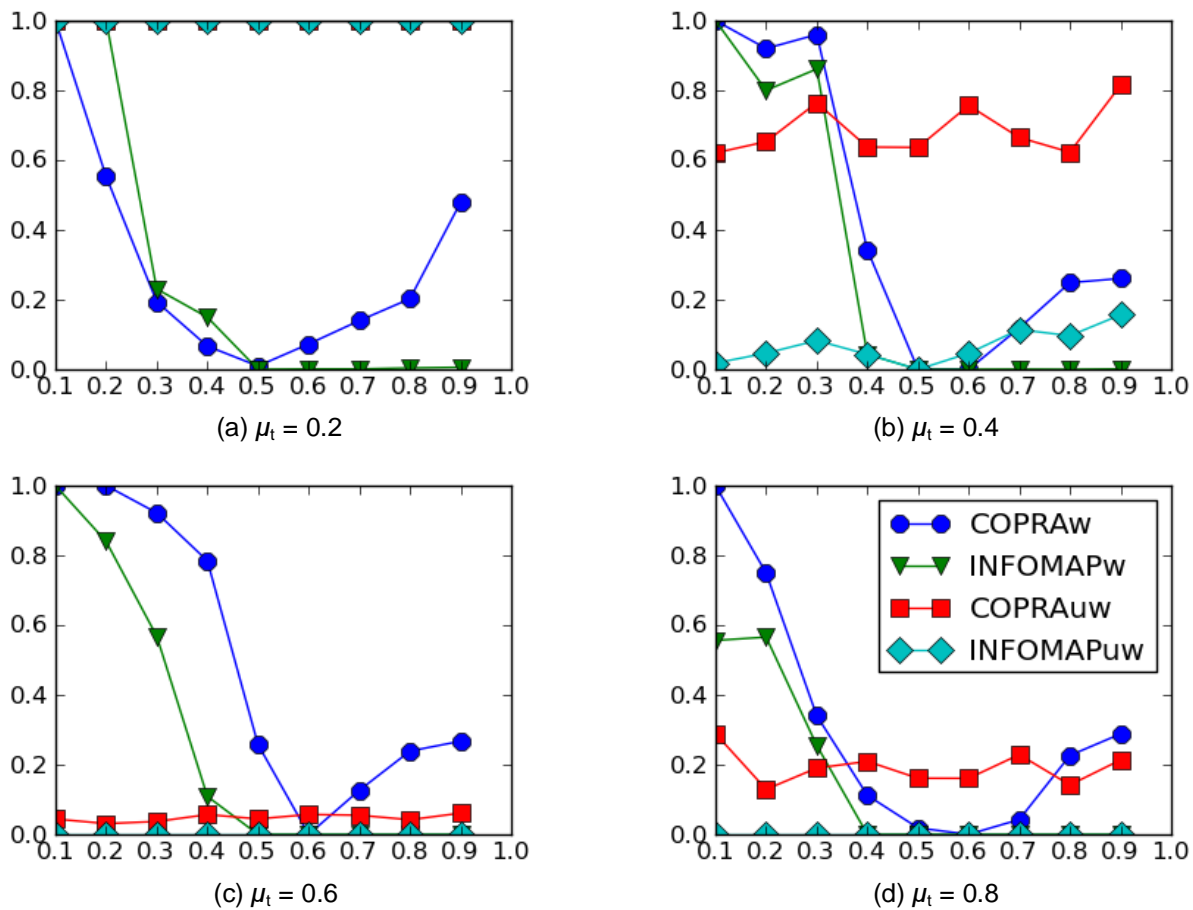

(a) $\mu_t = 0.2$

(b) $\mu_t = 0.4$

(c) $\mu_t = 0.6$

(d) $\mu_t = 0.8$

Figure 1: Mutual information scores (y-axis) as $\mu w$ (x-axis) changes. Each subplot shows a different fixed value for $\mu t$. The values of the other parameters were fixed: $N$=100, $k$=25, $\tau 1$=2, $\tau 2$=1, $\beta$=1.5

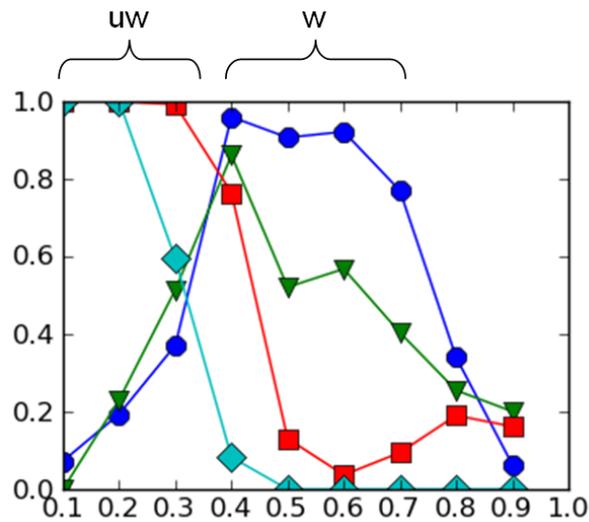

Figure 2: Mutual information scores for the weighted (w) and unweighted (uw) algorithms as $\mu t$ is varied. The value of $\mu w$ is fixed at 0.3. It is noticeable that the two classes of algorithm perform for complimentary settings of $\mu t$.

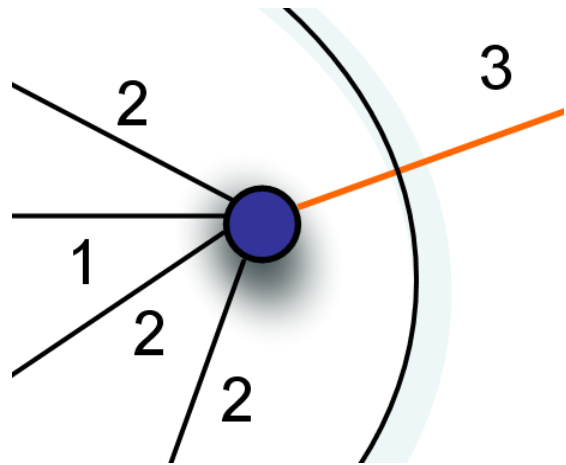

Figure 3: An example node with links and weights from a network with $\mu t = 0.2$ and $\mu w = 0.3$. As a result the single inter-community link (orange) receives a higher weight relative to the intra-community links.

The effect of this is that there are regions of the problem space, parameterised by community mixing proportions, in which a weighted algorithm will outperform an unweighted one and vice versa. This can be seen in Figure 2 where the two regions are labelled w (weighted) and uw (unweighted). This result indicates that a choice can be made, based on the community structure, as to the class of community detection algorithm.

In order to take advantage of this information and select the best class of algorithm for a given network, some knowledge of the underlying community structure is required. It may be possible to make some assumption about the communities that are sought after based on some knowledge of the specific domain. In most community detection problems however, this information about the community structure is unknown.

## 4  Parameter Estimation

In order to use the information from the previous section, it is required to know the values of the mixing parameters of the communities. Without knowledge of the communities (i.e. prior to community detection) it is not possible to evaluate these parameters. In this section it will be shown how parameters of the *observable* network can be mapped to these community parameters and how these values can be used to build a classifier to determine the class of community detection most suitable for the given network.

### 4.1 Observable Parameters

There are a range of metrics associated with describing network topology: degree distribution, average diameter, and centrality measures are a few of them. The problem here is that a parameter is required which describes the way that the community structures interact, without explicitly knowing the community structures.

To approach this, the node measure called clustering coefficient [25] is considered. This is defined as:

$$C_v^{(uw)} = \frac{\sum_{i,j \in N_v} e_{ij}}{k_v(k_v - 1)/2},\qquad(1)$$

where the local clustering coefficient, $C_v^{(uw)}$, represents the proportion of the neighbours, $N_v$, of node $v$ which are connected (i.e. edge $e_{ij}$=1 if there is a link between neighbouring nodes $i$ and $j$) out of the possible connections between its neighbours, $k_v(k_v\text{-}1)/2$. It was found that the mean value of the local clustering coefficient, taken over all the nodes in the network, showed a strong correlation with the topological mixing parameter, $\mu t$ (Figure 4a). This suggests that the mean clustering coefficient could be used to estimate this mixing parameter. If the mean clustering coefficient could be used to estimate the topological mixing then it follows that a weighted extension to this may yield information about the weighted mixing parameter (Equation 2).

$$C_v^{(w)} = \frac{\sum_{i,j \in N_v}(w_{vi} + w_{vj})e_{ij}}{\sum_{i \in N_v} w_{vi}(k_v - 1)},\qquad(2)$$

where $w_{vi}$ is the weight associated with the link between nodes $v$ and $i$. The mean of this value over the network was found to correlate with $\mu w$ (Figure 4b). The results in Figure 4 suggest that the mixing parameters can be estimated from observed network characteristics without knowledge of the community structure.

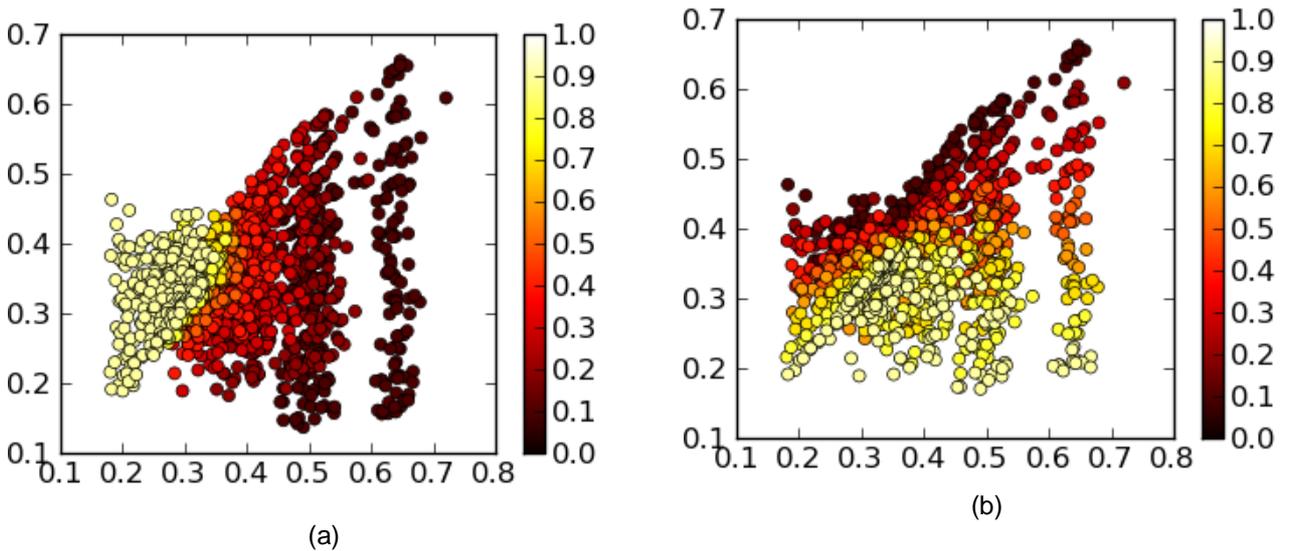

Figure 4: Scatter plots of the unweighted (x-axis) and weighted (y-axis) mean local clustering coefficient, (a) shows the value of the topology mixing parameter, $\mu t$. Similarly (b) shows the weight mixing parameter, $\mu w$.

The reason for this can be explained by considering the general principle of a community; that nodes within a community are more likely to be connected compared to overall probability of connection due to the sparse nature of the network. Hence, if two neighbours are within the same community, it is reasonable to expect them to be connected. However, if neighbours are not in the same community it is more likely that they are not connected. Based on this reasoning, the local clustering coefficient is an estimate of the individual node's mixing parameter, which averaged over the network yields a global estimate.

## 4.2 Algorithm Classification using SVM

The results of the previous section suggest that it is possible to estimate the mixing parameters of the communities. Now returning to the reason why it may be useful to estimate these parameters, i.e. to determine the class of algorithm, it is suggested that rather than estimate the mixing parameters and in turn predict the algorithm class, it may be more useful to use the clustering coefficients to directly predict the algorithm class. Figure 5 shows similar plots as Figure 4, but with the colour indicating the performance for the different algorithms. It can be seen that the weighted algorithms have a distinctly different performance pattern to the unweighted ones.

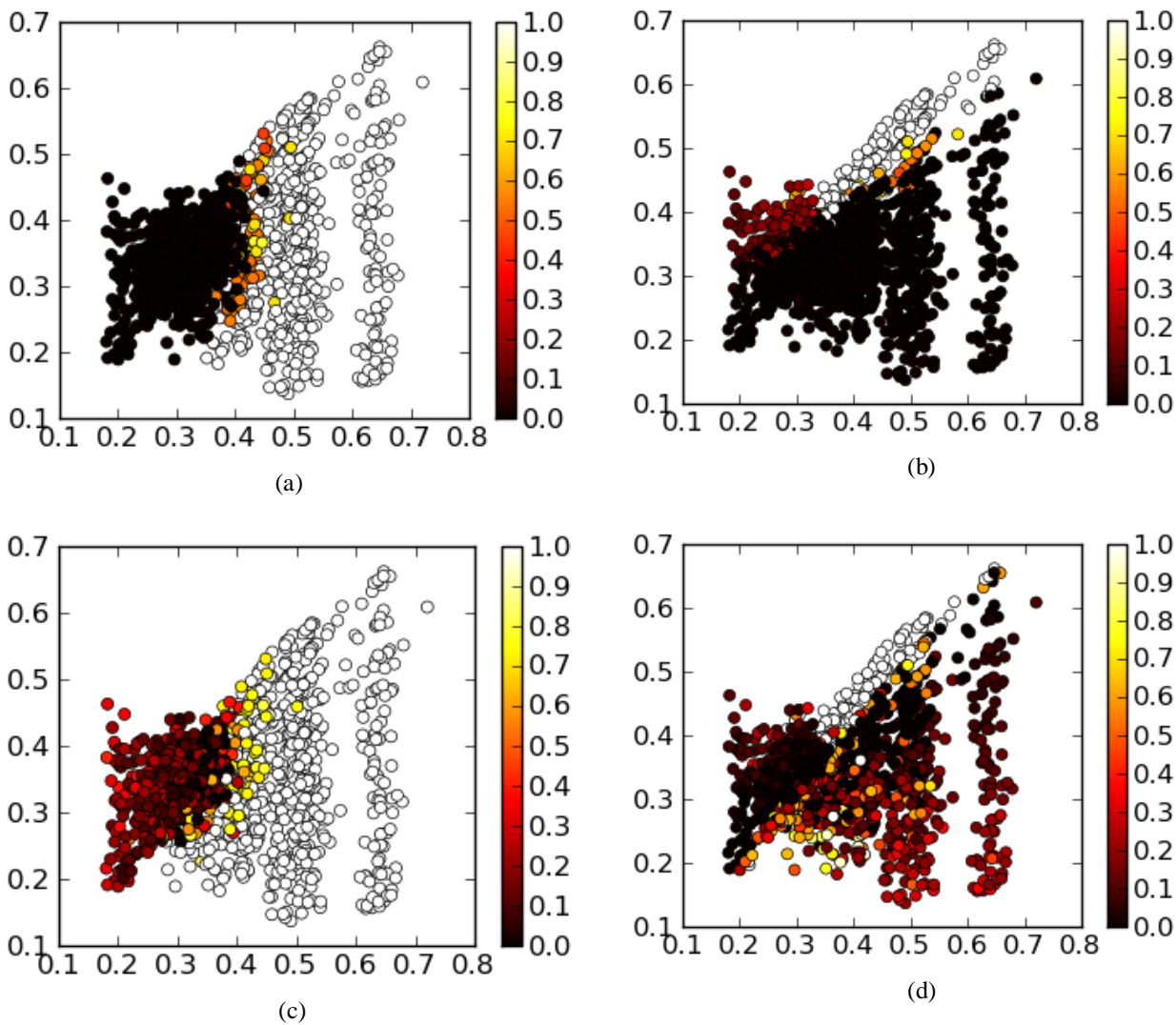

Figure 5: Clustering coefficients scatter plots with colours showing the mutual information score for (a) unweighted infomap, (b) weighted infomap, (c) unweighted COPRA and (d) weighted COPRA.

In order to confirm that these observable parameters can effectively predict the algorithm class, a simple classifier was built using linear support vector machines (SVM) [20]. To do this, each of the networks were assigned a class {weighted, unweighted, none} based on the class of algorithm which performed best in terms of its mutual information score. A class of "none" was assigned to any network where the mutual information score for the best performing algorithm was below some threshold. The reasoning for this is that for low performance values the output is not meaningful and therefore the choice of algorithm is irrelevant. As SVMs are restricted to two classes, three classifiers were trained (weighted vs. unweighted, weighted vs. none, unweighted vs. none) and the predicted class obtained by using a voting scheme over the three outputs. The results are discussed in the next section.

## 5   Results

A linear SVM was trained on 1790 networks taking the unweighted and weighted mean clustering coefficients as inputs. The "none" class was defined as networks for which the maximum mutual information score was below 0.6. The output classes for the test set (448 networks) are displayed in Figure 6. This can be compared to the true class labels in Figure 7. The overall performance on the test set was 83.9%. A confusion matrix of the test set performance is shown in Table 1.

To confirm these results, Figure 8 shows the mean performance, according to mutual information, when selecting the algorithm class using this classifier. This is compared against the performance of the best weighted algorithm and the best unweighted algorithm. From these graphs it can be seen that the classifier is able to select an appropriate class of algorithm such that it can achieve near optimum performance, constrained by the algorithms considered.

Table 1: Classifier Confusion Matrix

|  |  | Predicted Class | | |
|---|---|---|---|---|
|  |  | Weighted | Unweighted | None |
| True Class | Weighted | 42 | 3 | 36 |
|  | Unweighted | 4 | 125 | 18 |
|  | None | 6 | 5 | 209 |

From these results it can be seen that even with a simple classifier it is possible to obtain accurate predictions for the best class of community detection based on properties of the network alone.

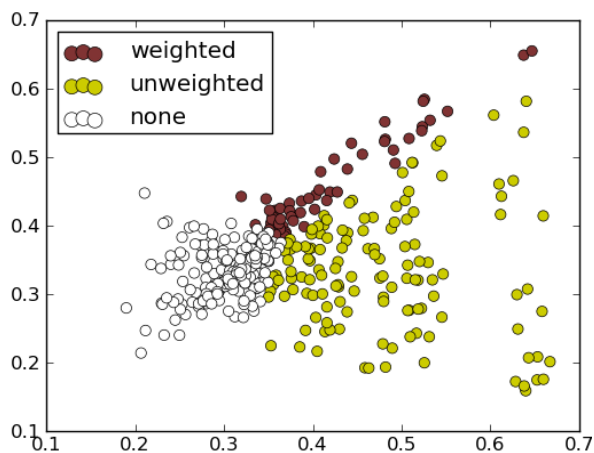

Figure 6: The predicted classification of the networks in the test set using a linear SVM.

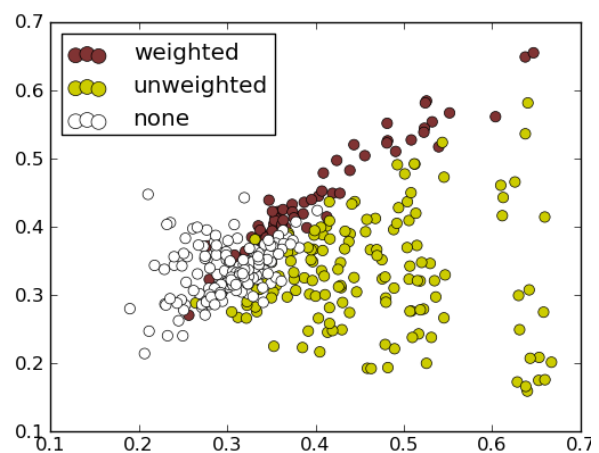

Figure 7: The true classification of the networks in the test set.

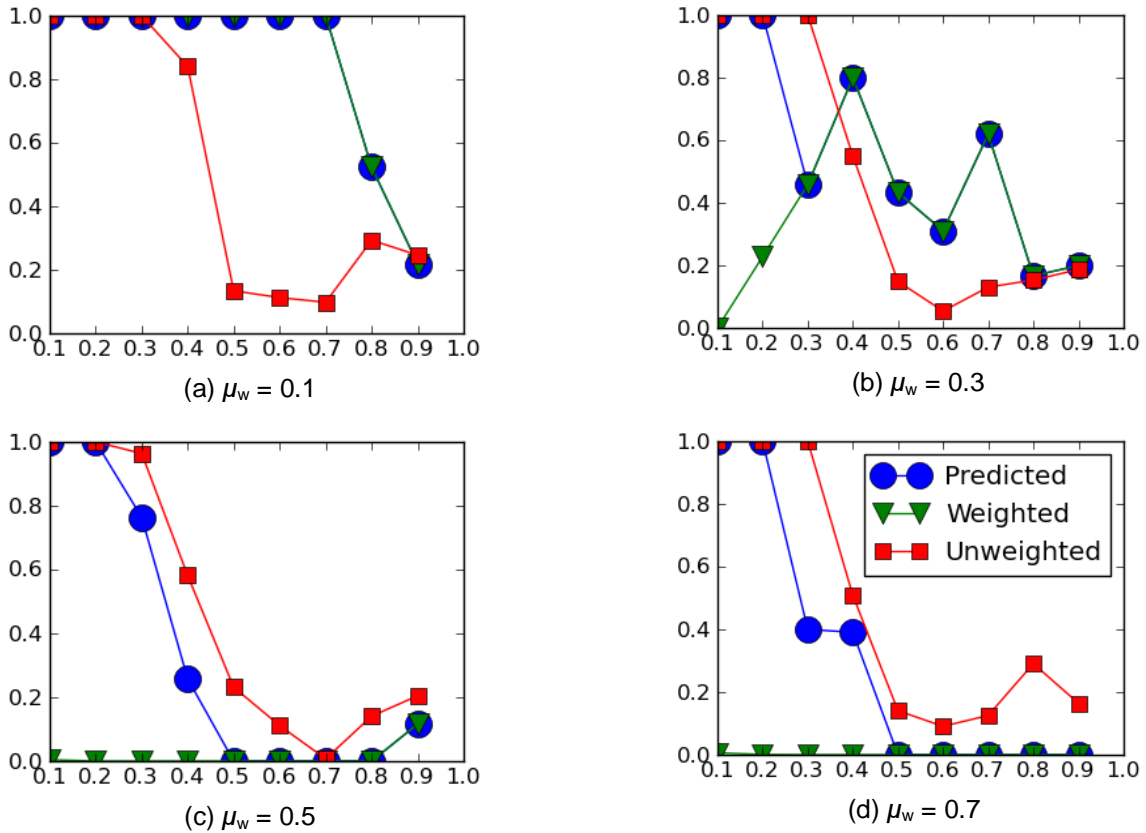

Figure 8: Mutual information scores for when algorithm class is selected by the classifier compared to the individual algorithm classes. Each graph shows the performance (y-axis) as $\mu t$ (x-axis) is varied for different $\mu w$ values.

## 6 Conclusion

To the best of the author's knowledge, no previous work has explored the problem of choosing an appropriate community detection algorithm based on the underlying structural properties. This work has presented community detection algorithms as examples of two classes of algorithm; weighted or unweighted. It is demonstrated that for different types of network and community structure, the class of algorithm has an effect on the performance. Furthermore it has been shown that it is possible to choose the algorithm class based only on the observed network parameters without prior knowledge of the community structure or assignment.

The algorithm selection in this work is highly constrained both in terms of the space of possible network and community structures and classes of algorithms considered. Future work will reduce these constraints by considering a wider range of input networks and a more comprehensive set of algorithms and classes.

## Acknowledgements

This work was undertaken as part of an Engineering Doctorate at the Computer Science department at University College London. UK Patent Pending No.: 1004376.8, March 2010